\documentclass[sigplan,screen]{acmart}
\usepackage[utf8]{inputenc}
\usepackage{graphicx}
\usepackage{listings}
\usepackage{cleveref}
\setcopyright{none}
\lstnewenvironment{dafny}{\lstset{language=dafny,basicstyle=\ttfamily\footnotesize}}{}
\usepackage{formaljitterbug}
\usepackage{algorithm,algpseudocode}
\renewcommand\footnotetextcopyrightpermission[1]{}
\acmConference[Dafny Workshop]{}{January 19, 2025}{Denver, Colorado, US}
\PassOptionsToPackage{hyphens}{url}\usepackage{hyperref}
\title{Verifying the Fisher-Yates Shuffle Algorithm in Dafny}

\author{Stefan Zetzsche}
\affiliation{
  \institution{Amazon Web Services}
  \country{United Kingdom}
}
\email{stefanze@amazon.co.uk}
\authornote{Corresponding author}

\author{Tancrede Lepoint}
\affiliation{
  \institution{Amazon Web Services}
  \country{United States}
}
\email{tlepoint@amazon.com}

\author{Jean-Baptiste Tristan}
\affiliation{
  \institution{Amazon Web Services}
  \country{United States}
}
\email{trjohnb@amazon.com}

\author{Mikael Mayer}
\affiliation{
  \institution{Amazon Web Services}
  \country{United States}
}
\email{mimayere@amazon.com}

\begin{abstract}
The Fisher-Yates shuffle is a well-known algorithm for shuffling a finite sequence, such that every permutation is equally likely. Despite its simplicity, it is prone to implementation errors that can introduce bias into the generated permutations. We verify its correctness in Dafny as follows. First, we define a functional model that operates on sequences and streams of random bits. Second, we establish that the functional model has the desired distribution. Third, we define an executable imperative implementation that operates on arrays and prove it equivalent to the functional model. The approach may serve as a blueprint for the verification of more complex algorithms.
\end{abstract}

\newcommand{\nat}{\mathbb{N}}

\begin{document}

\maketitle
\section{Introduction}

In this section, we give an introduction to the Fisher-Yates shuffle, the subject of format-preserving tokenization, and our larger approach for formalizing randomized algorithms.

\subsection{The Fisher-Yates Shuffle}

The Fisher-Yates shuffle is a well-known algorithm for shuffling a finite sequence, such that every permutation is equally likely. Its earliest form was described by Fisher and Yates \cite{fisher1938statistical}. Subsequent work by Knuth improved on its complexity and allowed for in-place shuffling \cite{knuth1969art}:

\begin{algorithm}[H] 
\caption{Fisher-Yates shuffle}
\label{alg:loop}
\begin{algorithmic}
	\For{$i \gets 0$ to $|a|-1$}                    
        \State {$j$ $\gets$ {random integer such that $i \leq j < |a|$}}
        \State {swap $a\lbrack i \rbrack$ and $a\lbrack j \rbrack$}
    \EndFor
\end{algorithmic}
\end{algorithm}

Despite its simplicity, the Fisher-Yates algorithm is prone to implementation errors that can introduce bias into the generated permutations~\cite{wikipedia-bias}. A simple off-by-one error in the bounds of $j$ turns it into Sattolo's algorithm, which only produces permutations with a single cycle. Such implementation flaws can have adverse effects, like skewing online lotteries or poker \cite{poker}. Formal verification can provide high assurance that the implementation is correct and unbiased. 

\subsection{Format-Preserving Tokenization}

The application of the Fisher-Yates shuffle we are particularly interested in lies in format-preserving tokenization. 

Format-preserving tokenization is a data protection technique that substitutes cleartext values with \emph{tokens} that retain original format and structure~\cite{ansi}. For example, a social security number \texttt{123-45-6789} could be tokenized into \texttt{481-01-7494}. The standard ANSI X9.119-2~\cite{ansi} outlines two main approaches: random tokens with a mapping table, or encryption-based tokens.  Random tokens work well for large inputs but risk collisions for small inputs. Encryption-based tokens avoid collisions but the NIST standard requires $r^\ell \geq 1,000,000$ where $r$ is the radix (i.e., number of characters in the alphabet) and $\ell$ is the length, limiting its use for short values like ZIP codes ($r=10, \ell=5$). This is where the Fisher-Yates shuffle becomes useful.

For short inputs, i.e. when $r^\ell < 1,000,000$, the full truth table representing the encryption can be generated with a random shuffler. Namely, we generate all the $r^\ell$ possible input values using a lexicographical order and then use the Fisher-Yates shuffle to create a random permutation of this array: this fully defines the format-preserving encryption function. To ensure that the tokenization method is secure, it is critical that the shuffle is uniformly random.

\subsection{Randomized Algorithms}

When talking about randomized algorithms, one can distinguish between two types: Las-Vegas and Monte-Carlo.

Las-Vegas algorithms always give the correct result and just use randomness internally.
An example is the randomized quicksort algorithm, which randomly chooses a pivot element to improve its running time on average.
The correctness specification of such algorithms does not mention randomness and is relatively easy to reason about in Dafny because one can treat randomness as nondeterminism.

Monte-Carlo algorithms, on the other hand, output random results and the correctness property concerns the distribution of the outputs.
An example is the Miller-Rabin primality test, which checks whether a property that is known to hold for prime numbers is satisfied by a given number. The test is a good indicator of an input being prime, but it can yield a wrong result with a very small probability.
Since Dafny does not have built-in support for probabilities, it is an interesting question how to best approach the verification of such probabilistic correctness properties.

The Fisher-Yates shuffle algorithm is of the Monte-Carlo type: it is correct if its outputs are uniformly distributed.

\subsection{Formalizing Randomness}

The simplest non-trivial random primitive is a fair independent coin flip yielding heads/tails, or true/false.
All (discrete) randomness can be reduced to such independent fair coin flips, which allows us to model a source of randomness as a countably infinite stream of independent and identically distributed fair random bits of type $\{0, 1\}^\nat$, that is, in Dafny
\begin{dafny}
type Bitstream = nat -> bool
\end{dafny}
as suggested by Joe Hurd in his PhD dissertation \cite{Hurd03}.
Under this view, sampling from a probability distribution reduces to transforming bitstreams to values in the domain $V$ of the desired distribution $D$.
To prove such a transformation $T: \{0, 1\}^\nat \to V$ correct, one has to show that for all samples $x \in V$, it holds $\mu(T^{-1}(\{x\})) = \Pr[X = x]$ where $X$ is a random variable with distribution $D$ and $\mu$ is a probability measure on bitstreams.
That is, the measure of the set of bitstreams that are transformed to the sample $x \in V$ should be equal to the probability of $x$ according to the distribution $D$.
For example, a coin flip can be expressed as:
\begin{dafny}
function Coin'(s: Bitstream): bool { 
  s(0) 
}
\end{dafny}
Under above view, its correctness translates to the equalities $\mu(\lbrace s \in \{0, 1\}^\nat \mid s(0) = b \rbrace) = 0.5$, where $b \in \lbrace 0, 1 \rbrace$. 

\subsection{An Overview of Our Approach}

Our approach for implementing and proving correct the Fisher-Yates shuffle consists of three parts:
\begin{itemize}
    \item a functional model that operates on sequences by using the bitstream transformer approach
    \item a correctness proof for the functional model, establishing that it has the desired distribution
    \item an executable imperative implementation that operates on arrays by sampling from an external uniform distribution and is proven equivalent to the functional model
\end{itemize}

\section{A Functional Model}

In this section, we will introduce a functional model of the Fisher-Yates shuffle.
\begin{figure}
    \centering
\begin{dafny}
datatype Result<T> = Result(value: T, rest: Bitstream)

type Hurd<T> = Bitstream -> Result<T>

function Return<T>(x: T): Hurd<T> {
  (s: Bitstream) => Result(x, s)
}

function Bind<S, T>(h: Hurd<S>, f: S -> Hurd<T>): Hurd<T> {
  (s: Bitstream) =>
    var (x, s') := h(s).Extract();
    f(x)(s')
}
\end{dafny}
\vspace{-0.5cm}
    \caption{The Hurd probability monad.}
    \label{fig:hurd}
\end{figure}

\subsection{The Hurd Monad}

As mentioned previously, we model the sampling from a probability distribution as the transformation of bitstreams to values in the domain of the distribution. To enable compositionality, we need to additionally return the stream of those bits that haven't been used during the transformation. As realized by Hurd \cite{Hurd03}, this access to randomness can be formalized as the state monad of type \texttt{Bitstream} (\Cref{fig:hurd}), which we call the \emph{Hurd monad}. Using it, we can rewrite the above \lstinline[language=dafny, basicstyle=\small\ttfamily]{Coin'} function as follows:
\begin{dafny}
function Coin(): Hurd<bool> {
  (s: Bitstream) =>
    Result(s(0), (n: nat) -> s(n + 1))
}
\end{dafny}
The \texttt{Coin} function returns a transformation that decomposes a stream into its first bit and the stream that starts at the second bit.
More generally, a generic function \texttt{Sample} that takes an input of type $\texttt{S}$ and returns a sample of type $\texttt{T}$, can be seen as returning a value of type \texttt{Hurd<T>}:
\begin{dafny}
function Sample<S,T>(x: S): Hurd<T>
\end{dafny}
The composition of such functions can be defined\footnote{The composition of \texttt{f: S -> Hurd<T>} and \texttt{g: T -> Hurd<U>} is defined as \texttt{(x: S) => Bind(f(x), g): S -> Hurd<U>}. The unit of the composition is \texttt{Return: T -> Hurd<T>}.} in terms of $\texttt{Bind}$ and is guaranteed to be e.g. associative. By combining the monadic primitives one can relatively easily construct more complex samplers that use random bits not more than once. For example, in previous work it was shown that \texttt{Coin} can be lifted to a function \texttt{Sample} that can be proven to correctly sample from the uniform distribution. In the present work we decided, for simplicity, to instead draw the line of axiomatization a bit higher, and assume the existence of such a function (\Cref{fig:sample}). Note that while related work in Lean since has found that the modelling of samplers of more complex distribution is easier done with a monad other than the Hurd monad \cite{Tristan_SampCert_Verified_2024}, the choice remains appropriate for modelling the Fisher-Yates shuffle, in our opinion.

\subsection{A Probability Space on Bitstreams}

To state the correctness of a sampler, we first need to introduce a probability measure on bitstreams. Even though we axiomatize the existence of such a probability space, its existence can in fact be proven \cite{Hurd03}. We decided to skip on such a proof as it involves the formalization of advanced mathematics and has already been done in Lean \cite{Tristan_SampCert_Verified_2024}. We begin by defining the set of measurable subsets of bitstreams:
 \begin{dafny}
ghost const eventSpace: iset<iset<Bitstream>>	
\end{dafny}
The probability measure assigning a value between $0$ and $1$ to sets of bitstreams is given by:
\begin{dafny}
ghost const prob: iset<Bitstream> -> real	
\end{dafny}
Finally, the predicate \texttt{IsProbability} (whose definition is not shown) asserts that \texttt{eventSpace} is a $\sigma$-algebra and that \texttt{prob} is a probability measure on that measurable space:
\begin{dafny}
lemma {:axiom} ProbIsProbabilityMeasure()
  ensures IsProbability(eventSpace, prob)
\end{dafny}

\begin{figure}
\begin{dafny}
ghost function {:axiom} Sample(n: nat): (h: Hurd<nat>)
  requires 0 < n
  ensures IsIndepFunction(h)
  ensures IsMeasurePreserving(eventSpace, prob, eventSpace,
                              prob, s => h(s).rest)
  ensures forall s :: 0 <= h(s).value < n
  ensures forall i | 0 <= i < n ::
              var e := iset s | h(s).value == i;
              && e in eventSpace
              && prob(e) == 1.0 / (n as real)
\end{dafny}
\vspace{-0.4cm}
\caption{Axiomatizing a sampler from the uniform distribution.}
\label{fig:sample}
\end{figure}

Thanks to this machinery, we can state that \texttt{Sample} returns values that are uniformly distributed (\Cref{fig:sample}).

It is straightforward to lift \texttt{Sample}, which for any positive integer \texttt{n} returns a value between \texttt{0} (inclusive) and \texttt{n} (exclusive), to a function \texttt{IntervalSample} that for any two integers \texttt{a} and \texttt{b} returns a value the between \texttt{a} (inclusive) and \texttt{b} (exclusive):
\begin{dafny}
ghost function IntervalSample(a: int, b: int): Hurd<int>
  requires a < b
{
  Map(Sample(b - a), x => a + x)
}	
\end{dafny}

Here \texttt{Map} denotes as usual a higher-order function that for any \texttt{f: S -> T} is of type \texttt{Map(-, f): Hurd<S> -> Hurd<T>} and satisfies \texttt{Map(h, f)(s).value == f(h(s).value)} and \texttt{Map(h, f)(s).rest == h(s).rest}.

One can show that the correctness of \texttt{Sample} implies the correctness of \texttt{IntervalSample}. That is, we can prove the following statement:
\begin{dafny}
lemma IntervalSampleCorrectness(a: int, b: int, i: int)
  requires a <= i < b
  ensures
    var e := iset s | IntervalSample(a, b)(s).value == i;
    && e in Rand.eventSpace
    && Rand.prob(e) == (1.0 / ((b-a) as real))
\end{dafny}

\subsection{Compositionality and Independence}

\begin{figure}
    \centering
\begin{dafny}
ghost predicate IsIndepFunction<A(!new)>(h: Hurd<A>) {
  forall A: iset<A>, E: iset<Bitstream> | E in eventSpace :: 
    IsIndepFunctionCondition(h, A, E)
}

ghost predicate IsIndepFunctionCondition<A(!new)>
(h: Hurd<A>, A: iset<A>, E: iset<Bitstream>) {
  AreIndepEvents(
    eventSpace,
    prob,
    BitstreamsWithValueIn(h, A),
    BitstreamsWithRestIn(h, E))
}

predicate AreIndepEvents<T>
(eventSpace: iset<iset<T>>, prob: iset<T> -> real, 
e1: iset<T>, e2: iset<T>) {
    && (e1 in eventSpace)
    && (e2 in eventSpace)
    && (prob(e1 * e2) == prob(e1) * prob(e2))
}
\end{dafny}
\vspace{-0.4cm}
    \caption{The (weak) functional independence property.}
    \label{fig:independence}
\end{figure}

Restricting oneself to the monadic primitives \texttt{Return} and \texttt{Bind}, the coin flip \texttt{Coin}, and functions defined exclusively in terms of these constructs, has the advantage of ensuring that the computed value and the unused rest of a bitstream are probabilistically independent.
This is desirable because it allows compositional reasoning: properties of sub-samplers still hold if used in the context of other samplers.
One way to \emph{violate} independence would be to define the following:
\begin{dafny}
function BadCoin(): Hurd<bool> {
  (s: Bitstream) => Result(s(0), s)
}
\end{dafny}
The above function still returns the first bit of a bitstream, but doesn't shift the latter by one.
In consequence, the composition of \texttt{BadCoin} and \texttt{Coin}
returns a pair of perfectly correlated coin flips instead of independent coin flips.

To capture the subset of those monadic computations for which \texttt{value} and \texttt{rest} are probabilistic independent, Hurd introduced the notion of \emph{strong independence} \cite{Hurd03}. One can show that the function \texttt{BadCoin} violates this notion.
The formal definition is quite technical (for example, it ensures that all involved sets are measurable), so we refer to \cite[Def. 35]{Hurd03} for the details.  In fact, relevant for our purposes is mostly a strictly weaker variation of it: \emph{weak (functional) independence} \cite[Def. 33]{Hurd03}. We formalize the latter as a predicate \texttt{IsIndepFunction} (\Cref{fig:independence}).
For a computation \texttt{h: Hurd<T>}, the property \texttt{IsIndepFunction(h)} ensures that the functions \texttt{s => h(s).value} and \texttt{s => h(s).rest} are measurable and that for any \texttt{T}-event \texttt{A} and \texttt{Bitstream}-event \texttt{E}, the events \texttt{(iset s | h(s).value in A)} and \texttt{(iset s | h(s).rest in E)} are probabilistically independent.

 Hurd has shown that one can construct a uniform sampler on top of \texttt{Coin} that satisfies strong independence, i.e. in particular weak (functional) independence \cite[Eq. 4.10]{Hurd03}. Since here we formalize such a sampler via the function \texttt{Sample}, it is sound to assume \texttt{IsIndepFunction(Sample(n))} for all \texttt{n: nat} (\Cref{fig:sample}). In consequence, we can prove the following statement:
\begin{dafny}
lemma IntervalSampleIsIndepFunction(a: int, b: int)
  requires a < b
  ensures IsIndepFunction(IntervalSample(a, b))	
\end{dafny} 
 Finally, we assume that the function \texttt{s => Sample(n)(s).rest} is \emph{measure-preserving} (\Cref{fig:measure_preserving}). Again, it is possible to lift the property from \texttt{Sample} to \texttt{IntervalSample}, that is, we can prove the following statement:
 \begin{dafny}
 lemma IntervalSampleIsMeasurePreserving(a: int, b: int)
  requires a < b
  ensures 
    IsMeasurePreserving(eventSpace, prob, eventSpace, prob, 
                        s => IntervalSample(a, b)(s).rest)	
 \end{dafny}

\begin{figure}
\begin{dafny}
ghost predicate IsMeasurePreserving<S(!new),T>
(eventSpaceS: iset<iset<S>>, measureS: iset<S> -> real, 
 eventSpaceT: iset<iset<T>>, measureT: iset<T> -> real, 
 f: S -> T) {
  && IsMeasurable(eventSpaceS, eventSpaceT, f)
  && forall e | e in eventSpaceT :: 
    measureS(PreImage(f, e)) == measureT(e)
}

ghost predicate IsMeasurable<S(!new),T>
(eventSpaceS: iset<iset<S>>, eventSpaceT: iset<iset<T>>, 
 f: S -> T) {
 forall e | e in eventSpaceT :: PreImage(f,e) in eventSpaceS
}

ghost function PreImage<S(!new),T>
(f: S -> T, e: iset<T>): iset<S> {
  (iset s | f(s) in e)
}
\end{dafny}
\vspace{-0.5cm}
\caption{The measure-preserving property.}
\label{fig:measure_preserving}
\end{figure}

\subsection{A Recursive Model of Fisher-Yates}

\begin{figure}
\begin{dafny}
ghost function Shuffle<T>(xs:seq<T>,i:nat := 0): Hurd<seq<T>>
  requires i <= |xs|
{
  (s: Bitstream) => ShuffleCurried(xs, s, i)
}

ghost function ShuffleCurried<T>
(xs: seq<T>, s: Bitstream, i: nat := 0): Result<seq<T>>
  requires i <= |xs|
  decreases |xs| - i
{
  if |xs| > 1 + i then
    var (j, s') := IntervalSample(i, |xs|)(s).Extract();
    var ys := Swap(xs, i, j);
    ShuffleCurried(ys, s', i + 1)
  else
    Return(xs)(s)
}	
\end{dafny}	
\vspace{-0.3cm}
\caption{A functional model of the Fisher-Yates shuffle.}
\label{fig:fisher_yates_functional}
\end{figure}

In \Cref{fig:fisher_yates_functional}, we present \texttt{Shuffle}, a purely functional implementation of the Fisher-Yates algorithm that is of the generic type \texttt{seq<T> -> Hurd<seq<T>>}.
This recursive definition operates on sequences instead of arrays and samples from the function \texttt{IntervalSample} we introduced earlier.

\section{A Correctness Proof}

\label{sec:correctness-proof}

In this section, we will establish the correctness of the functional model \texttt{Shuffle}. 

\subsection{Specifying Correctness}   

Intuitively, we say that \texttt{Shuffle} is correct, if for any \texttt{xs: seq<T>} and permutations \texttt{p, q: seq<T>} of \texttt{xs} (that is, as multisets, the three are equal), it as likely that applying \texttt{Shuffle} to \texttt{xs} results in \texttt{p}, as it is that it results in \texttt{q}. That is, the probability that the outcome is \texttt{p} should be \texttt{1} over the number of permutations of \texttt{xs}. We restrict ourselves to the case where \texttt{xs} contains no duplicates, which implies that the number of permutations of \texttt{xs} is the factorial of its length. This restriction does not result in a loss of generality, because we can map an arbitrary sequence \texttt{xs} to the sequence \texttt{seq(|xs|, i requires 0 <= i < |xs| => (xs[i], i))} without duplicates. Formally, we specify the correctness of \texttt{Shuffle} as the following lemma, which we will prove next: 
\begin{dafny}
lemma Correctness<T(!new)>(xs: seq<T>, p: seq<T>)
  requires forall a, b | 0 <= a < b < |xs| :: xs[a] != xs[b]
  requires multiset(p) == multiset(xs)
  ensures
    var e := iset s | Shuffle(xs)(s).value == p;
    && e in eventSpace
    && prob(e) == 1.0 / (Factorial(|xs|) as real)	
\end{dafny}

\subsection{Proving Correctness}

Instead of attempting at a direct proof of \texttt{Correctness}, we establish a slightly more general variant of it:

\begin{dafny}
lemma CorrectnessGeneral<T(!new)>(xs:seq<T>, p:seq<T>, i:nat)
  decreases |xs| - i
  requires i <= |xs| && |xs| == |p|
  requires forall a, b | i <= a < b < |xs| :: xs[a] != xs[b]
  requires multiset(p[i..]) == multiset(xs[i..])
  ensures
    var e := iset s | Shuffle(xs, i)(s).value[i..] == p[i..]
    && e in eventSpace
    && prob(e) == 1.0 / (Factorial(|xs|-i) as real)
\end{dafny}

Clearly \texttt{Correctness(xs, p)} follows immediately from \texttt{CorrectnessGeneral(xs, p, 0)}. We divide the proof of \texttt{CorrectnessGeneral(xs, p, i)} into two parts. In the case \texttt{|xs[i..]| <= 1}, we call \texttt{CorrectnessGeneralLeqOne(xs, p, i)}, and otherwise \texttt{CorrectnessGeneralGreOne(xs, p, i)}. In both cases, we define \texttt{e := iset s | Shuffle(xs, i)(s).value[i..] == p[i..]}.
\begin{itemize}
	\item For the proof of \texttt{CorrectnessGeneralLeqOne(xs, p, i)}, we first note that \texttt{|xs| <= 1 + i}, since  \texttt{|xs[i..]|}  is equal to \texttt{|xs| - i}. By the definition of \texttt{Shuffle}, we thus get \texttt{Shuffle(xs, i)(s) == Return(xs)(s)}. From the latter, we deduce that the event \texttt{e} is equal to the full sample space, which consists of all possible bitstreams. Since \texttt{prob} is a probability measure, we find \texttt{prob(e) == 1}, which concludes the proof, as \texttt{Factorial(|xs|-i) == 1}. 
\item 
For the proof of \texttt{CorrectnessGeneralGreOne(xs, p, i)}, let \texttt{A := iset j | i <= j < |xs| \&\& xs[j] == p[i]}. Then one can show that there exists exactly one \texttt{j}, such that \texttt{A == iset\{j\} }. This allows us to define \texttt{ys := \texttt{Swap}(xs, i, j)} and \texttt{e' := iset s | Shuffle(ys, i+1)(s).value[i+1..] == p[i+1..]}. Let \texttt{h := IntervalSample(i, |xs|)}, then \texttt{e} is the intersection of \texttt{BitstreamsWithValueIn(h, A)} and  \texttt{BitstreamsWithRestIn(h, e')}. Since \texttt{h} satisfies weak functional independence, we can deduce that \texttt{prob(e)} is equal to the product of the probabilities \texttt{prob(BitstreamsWithValueIn(h, A))} and \\ \texttt{prob(BitstreamsWithRestIn(h, e'))}. It remains to compute both of the factors:
\begin{itemize}
\item For the first factor, we notice that the infinite set \texttt{BitstreamsWithValueIn(h, A)} is equal to \texttt{(iset s | IntervalSample(i, |xs|)(s).value == j)}. From the correctness of \texttt{IntervalSample} it thus follows that \texttt{prob(BitstreamsWithValueIn(h, A))} is \texttt{1.0} divided by \texttt{|xs|-i} as real.
\item For the second factor, we notice that the probability \texttt{prob(BitstreamsWithRestIn(h, e'))} is equal to \texttt{prob(e')}, since \texttt{h} is measure-preserving. We then deduce that \texttt{prob(e')} is \texttt{1.0} divided by the factorial \texttt{Factorial(|xs|-(i+1))} as real from i) \texttt{|xs| == |ys|} and calling \texttt{CorrectnessGeneralLeqOne(ys, p, i+1)}, if \texttt{|ys[i+1..]| <= 1}; and ii) from calling \texttt{CorrectnessGeneralGreOne(ys, p, i+1)} otherwise.
\end{itemize}
 Finally, we observe that \texttt{Factorial(|xs|-i)} is equal to the product of \texttt{|xs|-i} and \texttt{Factorial(|xs|-(i+1))} by the definition of \texttt{Factorial}.
\end{itemize}
\section{An Imperative Implementation}

In this section, we introduce an executable imperative implementation of the Fisher-Yates shuffle and prove it equivalent to the functional model.

\subsection{External Randomness}

A functional model like \texttt{Shuffle} is well suited for mathematical reasoning about its correctness properties.
In practice, however, a probabilistic program will use a real-world source of random bits rather than an abstract infinite stream of random bits.
Since the state of the underlying random bit generator will be modified during such a  sampling process, we use Dafny's \texttt{method} for an executable implementation. With the Dafny compiler and a target language's probabilistic primitives, the translation of a method to a variety of languages such as C\#, Python, and Java is possible.

At the heart of our integration with external randomness lies the following method that is part of a trait and is implemented by an external source of uniformly distributed random numbers:
\begin{dafny}  
method Sample(n: nat) returns (i: nat)
  modifies `s
  decreases *
  requires 0 < n
  ensures 0 <= i < n
  ensures Model.Sample(n)(old(s)) == Result(i, s)
\end{dafny}

We assume the external implementation to be correct, in the sense that it is an instance of the functional model \texttt{Sample}, which is contained in a module \texttt{Model}. The correspondence is formalized as \texttt{Model.Sample(n)(old(s)) == Result(i, s)}, where \texttt{s} refers to an internal ghost variable of type \texttt{Bitstream} that is used for verification purposes only. In principle, it is possible to make a weaker assumption by starting with an external source of coin flips instead of uniformly distributed random numbers. Making such an assumption leads to a smaller trusted code base, but makes less sense when optimizing for run-time efficiency and simplicity, as is the case here. 

By building on top of the externally implemented method \texttt{Sample} above, we are also able to define an imperative version of the functional model \texttt{IntervalSample}, which is contained in a module \texttt{Model}, as follows:

\begin{dafny}
method IntervalSample(a: int, b: int) returns (i: int)
  modifies `s
  decreases *
  requires a < b
  ensures a <= i < b
  ensures Model.IntervalSample(a, b)(old(s)) == Result(i, s)
{
  var j := Sample(b - a);
  assert Model.Sample(b-a)(old(s)) == Result(j,s);
  assert Model.IntervalSample(a,b)(old(s)) == Result(a+j,s);
  i := a + j;
}	
\end{dafny}

The equivalence between the functional and imperative implementation is again captured by the \texttt{ensures} clause \texttt{Model.IntervalSample(a, b)(old(s)) == Result(i, s)}. This time, we can prove its correctness (with \texttt{assert} statements), by utilizing some of the properties of \texttt{Sample}.

\subsection{An Executable Implementation Of Fisher-Yates}

\begin{figure}
    \centering
\begin{dafny}
method Shuffle<T>(a: array<T>)
 decreases *
 modifies `s, a
 ensures Model.Shuffle(old(a[..]))(old(s)) == Result(a[..],s)
{
  if a.Length > 1 {
    for i := 0 to a.Length - 1 {
      var j := IntervalSample(i, a.Length);
      Swap(a, i, j);
}}}
\end{dafny}
\vspace{-0.4cm}
    \caption{An imperative implementation of the Fisher-Yates shuffle. We omit loop invariants and assertions.}
        \label{fig:fisher_yates_imperative_simplified}
\end{figure}

In \Cref{fig:fisher_yates_imperative_simplified}, we present an executable implementation of the Fisher-Yates shuffle that utilizes the previously defined imperative version of \texttt{IntervalSample}. It is instructive to compare this implementation of \texttt{Shuffle} to its functional model in \Cref{fig:fisher_yates_functional}, in particular for the case in which \texttt{i} is \texttt{0}. Note that, among others, above version operates on arrays instead of sequences.
The equivalence between the functional model and its imperative implementation is formalized as the clause  \texttt{Model.Shuffle(old(a[..]))(old(s)) == Result(a[..], s)}. It essentially says that applying the executable implementation to an array before converting it to a sequence is the same as first converting the array to a sequence and then applying the functional model. For Dafny to establish the equivalence, we have to provide it with an appropriate for-loop invariant and assertions. In \Cref{fig:fisher_yates_imperative_simplified} we omit all such ghost code for readability. In reality, we keep track of three ghost variables, \texttt{prevI}, \texttt{prevASeq}, and \texttt{prevS}, which we set at the beginning of every loop iteration to \texttt{i}, \texttt{a[..]}, and \texttt{s}, respectively. The invariant then becomes \texttt{LoopInvariant(prevI, i, a, prevASeq, old(a[..]), old(s), prevS, s)}, where \texttt{LoopInvariant} is the ghost predicate defined by:

\begin{dafny}
ghost predicate LoopInvariant<T>(prevI: nat, i: nat, 
  a: array<T>, prevASeq: seq<T>, oldASeq: seq<T>, 
  oldS: Bitstream, prevS: Bitstream, s: Bitstream)
  reads a
{
  && prevI <= |prevASeq|
  && i <= a.Length - 1
  && (Model.Shuffle(oldASeq)(oldS) == 
      Model.Shuffle(prevASeq, prevI)(prevS))
  && (Model.Shuffle(prevASeq, prevI)(prevS) == 
      Model.Shuffle(a[..], i)(s))
}	
\end{dafny}

\section{Summary}

In this section, we conclude with some final remarks and discuss potential directions for the extension of the project.

\subsection{Assumptions}

We summarise all the assumptions that were made. First, we axiomatized the lemma \texttt{ProbIsProbabilityMeasure}, which states that \texttt{eventSpace} and \texttt{prob} form a probability space on the set of \texttt{Bitstreams}. Second, we axiomatized the existence of a function \texttt{Model.Sample} that correctly returns uniformly distributed random numbers (\Cref{fig:sample}). Previous work has shown that both assumptions are sound \cite{Zetzsche_Dafny-VMC_a_Library_2023,Hurd03}. Finally, we assume the existence of an (external) method \texttt{Sample} that implements \texttt{Model.Sample}. Here, we depend on the correctness of the probabilistic primitives in the languages Dafny compiles to.

\subsection{Presentation}

For a clear presentation of the underlying work, we at times allowed for a few minor simplifications. If it helped, entities were given a different identifier than in the actual implementation. To compile to Java's \texttt{int} instead of \texttt{BigInteger}, we actually used Dafny's bounded \texttt{int32} instead of the unbounded \texttt{int}. Surprisingly, there was almost no proof overhead caused by this complication.

\subsection{Future Work}

The development is part of a larger library of verified probabilistic samplers \cite{Zetzsche_Dafny-VMC_a_Library_2023} that for most parts has been moved to Lean for the proof development \cite{Tristan_SampCert_Verified_2024}. The Fisher-Yates shuffle algorithm is different to other parts of the library in the sense that it operates on the heap. At the moment, we don't have any plans for continuing the development of the library in Dafny.

\bibliographystyle{ACM-Reference-Format}
\bibliography{references}

\end{document}